# Quantum state magnification

Utilizing collective atomic interactions, we demonstrate quantum metrology without the need for detection sensitivity below the standard quantum limit.

O. Hosten[1], R. Krishnakumar[1], N. J. Engelsen[1], M. A. Kasevich[1*]

[1]Department of Physics, Stanford University, Stanford, California 94305, USA

*To whom correspondence should be addressed; E-mail: kasevich@stanford.edu

**Quantum metrology exploits entangled states of particles to improve sensing precision beyond the limit achievable with uncorrelated particles. All previous methods required detection noise levels below this standard quantum limit to realize the benefits of the intrinsic sensitivity provided by these states. Remarkably, a recent proposal has shown that, in principle, such low-noise detection is not a necessary requirement. Here, we experimentally demonstrate a widely applicable method for entanglement-enhanced measurements without low-noise detection. Using an intermediate magnification step, we perform squeezed state metrology 8 dB below the standard quantum limit with a detection system that has a noise floor 10 dB above the standard quantum limit. Beyond its conceptual significance, this method eases implementation complexity and is expected to find application in next generation quantum sensors.**

The prospect of using quantum entanglement to improve the precision of atomic and optical sensors has been a topic of discussion for more than two decades. Experiments aimed at achieving this objective have shown substantial improvements in the last half a decade. Exemplary work using atomic ensembles include preparation of spin-squeezed states *(1-12)*, Dicke states *(13-15)*, and other states with negative Wigner functions *(16)*. An assumption common to all this work is that low-noise detection methods are required to properly measure and utilize the prepared quantum states. In fact, detection noise has thus far been the bottleneck in the performance of these systems. To this end, there has been dedicated work on improving state-selective detection of atoms with both optical cavity aided measurements *(17, 18)* and fluorescence imaging *(19,20)*.

Here we describe the concept and the implementation of a quantum state magnification technique that relaxes stringent requirements in detection sensitivity for quantum metrology. This method is a generalization of a recent proposal for approaching the Heisenberg limit in measurement sensitivity without single-particle detection *(21)*. We demonstrate the method in an ensemble of half a million $^{87}$Rb atoms. As in a typical atomic sensor or clock, the goal is to measure a differential phase shift accumulated between a pair of quantum states during a time interval. To make this measurement, one first converts the phase shift into a population difference *(22, 12)*, and then precisely measures the population difference. The scheme we describe magnifies this population difference before the final detection. In our experiment, the atomic ensemble is first spin-squeezed using atomic interactions aided by an optical cavity, and then small rotations – to be sensed – are induced on the atomic state. These rotations are magnified up to 100 times by stretching the rotated states (Fig. 1A), again using cavity-aided interactions, and are finally detected via fluorescence imaging. Crucially, magnification allows for substantial reduction in the noise requirements for the final detection. While the method is demonstrated in an atom/cavity system, it is broadly applicable to any quantum system which has a suitable nonlinear interaction [see below, and Supplementary Materials *(23)*].

The collective state of an ensemble of $N$ two-level atoms – here the clock states of $^{87}$Rb – can be described using the language of a pseudo spin-$N/2$ system. The z-component of the spin, $J_z$, represents the population difference, and the orientation in the $J_x - J_y$ plane represents the phase difference between the two states. As these angular momentum components do not commute, both the population and the phase possess uncertainties. For a state with $\langle J_x \rangle \approx N/2$ the uncertainties satisfy $\Delta J_z \cdot \Delta J_y \geq N/4$, where $J_y$ is now identified with the phase of the ensemble. Coherent spin states (CSS) with noise $\Delta J_z = \Delta J_y = \sqrt{N}/2 \equiv \Delta_{CSS}$ establish the standard quantum limit (SQL) to minimum resolvable phase or population difference.

The magnification concept is illustrated in Figure 1A. The procedure starts with a mapping of $J_z$ onto $J_y$ (Fig. 1B) via a shearing interaction. A simple rotation of $J_y$ into $J_z$ follows to complete the sequence. The interaction leading to the mapping (shearing) generates a rotation of the state about the z-axis with the rotation rate depending on $J_z$, and is represented by the one-axis twisting Hamiltonian *(24)* $H = \hbar \chi J_z^2$, where $\chi$ is the shearing strength.

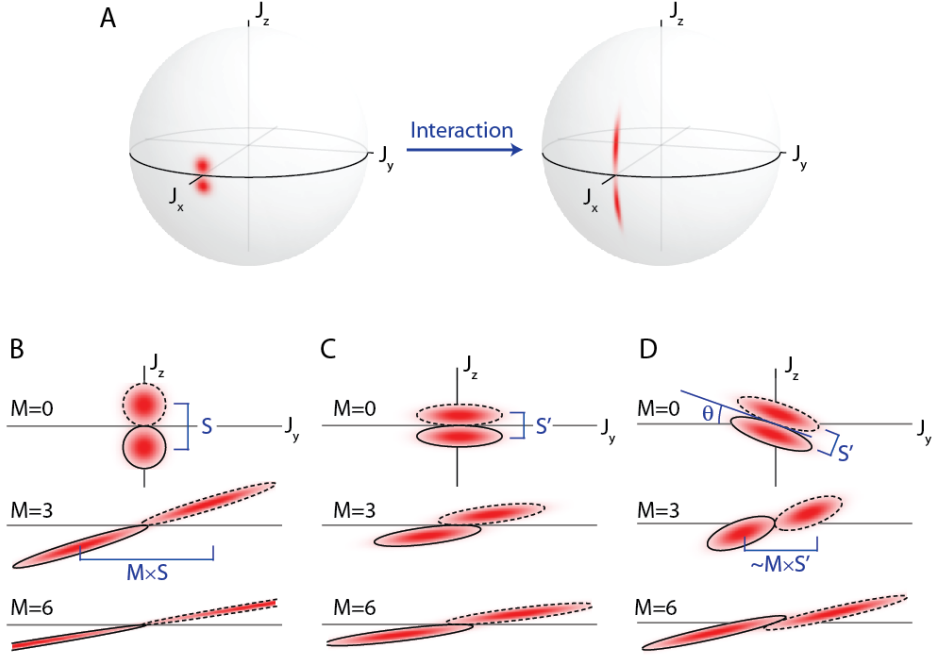

**Fig. 1. Conceptual description of quantum state magnification.** (**A**) Illustration of the state magnification protocol on the Bloch sphere. The Wigner quasi-probability distributions are shown for two separated initial CSSs (left), and after the states are magnified through collective interactions (right). Here this is shown with $N = 900$ atoms and a magnification of $M = 3$ for pictorial clarity. Experimentally we use up to $N = 5{\times}10^5$ and $M = 100$, permitting us to concentrate on a planar patch of the Bloch sphere. (**B-C**) Effect of the $J_z^2$ (shearing) interaction used for mapping $J_z$ onto $J_y$ for a pair of different initial states with separations $S$ and $S'{=}S/2$; each panel shows three different magnification factors. Note that a $\sim\pi/2$ on-axis rotation needs to follow to complete the protocol. (B) shows CSSs and (C) shows 6 dB squeezed states, together illustrating the requirement of larger magnifications to separate two initially squeezed states. (**D**) A small rotation $\theta$ before the shearing step is added, eliminating the requirement of larger magnifications for squeezed states by giving rise to a re-focusing of the $J_y$ noise. At an optimal magnification (here $M = 3$), the noise re-focusing scheme maps the initial $J_z$ onto $J_y$ with the SNR of the two initial states preserved.

The Heisenberg equations of motion for the vector operator $\mathbf{J}$ yields $\dot{\mathbf{J}} = \frac{1}{2}(\mathbf{\Omega}{\times}\mathbf{J} - \mathbf{J}{\times}\mathbf{\Omega})$, where the rotation vector $\mathbf{\Omega} = \hat{\mathbf{z}}\,2\chi J_z$ is also an operator. For our experimental parameters, the involved uncertainties will always be a small fraction of the Bloch sphere, so we can assume small angles and near-maximal initial x-polarization $J_x \approx J = N/2$ to restrict ourselves to a planar phase space (Fig. 1B-D). The equations then yield $J_z(t) = J_z(0)$ and $J_y(t) = J_y(0) + M J_z(0)$ with $M = N \int_0^t dt' \chi(t')$. Thus the initial $J_z$ is mapped onto $J_y$ with a magnification factor $M$. This is analogous to free expansion of a gas if one identifies $J_z$ with a particle's momentum and $J_y$ with its position.

We implement the one-axis twisting Hamiltonian through a dispersive interaction between the atoms and an optical cavity *(1)* (Fig. 2A). The underlying mechanism is a coupling between the intra-cavity power and atomic populations. The atom-cavity detuning is set such that the shift in the cavity resonance due to the atoms is proportional to $J_z$ (Fig. 2C). Thus $J_z$ sets the cavity-light detuning, which in turn sets the intra-cavity power (Fig. 2D), which in turn provides a $J_z$ dependent ac-Stark shift – hence the $J_z^2$ interaction. For our system the magnification parameter is $M = N \frac{\delta_c \delta_0}{\delta_0^2 + (\kappa/2)^2} \phi_{AC}$ *(23)* with $\delta_c = 5.5$ Hz, the cavity frequency shift per unit $J_z$; $\kappa = \{8.0, 10.4\}$ kHz, the cavity full-linewidth at $N = \{0, 5 \times 10^5\}$; $\delta_0$, the empty cavity-light detuning; and $\phi_{AC}$, the integrated phase shift of the clock states (precession on Bloch sphere) due to the interaction light. In reality, the decay of the cavity field results in back-action noise that is not taken into account in the simple Hamiltonian analysis above. However, these effects are insignificant in the parameter range we use for the magnification protocol, and can be ignored *(23)*.

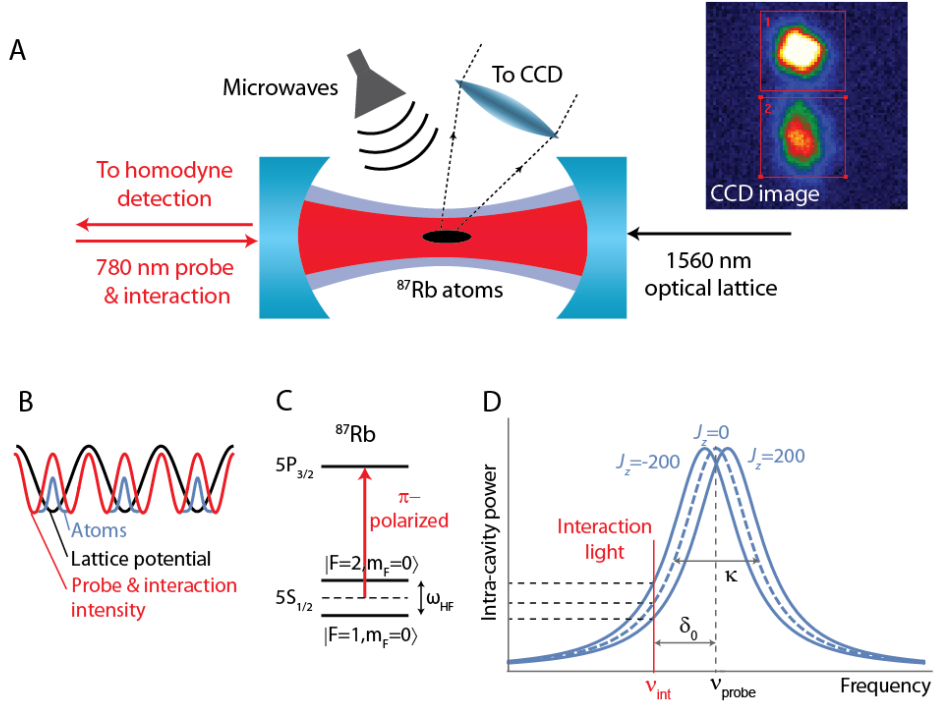

**Fig. 2. Experimental setup implementing state magnification.** (**A**) $^{87}$Rb atoms are trapped inside a 10.7 cm long high-finesse cavity using a 1560 nm cavity mode as a 1D optical lattice. A 780 nm mode is used to generate collective atomic interactions and to probe the cavity resonance frequency ($J_z$ measurements) by recording the phase of a reflected ~10 pW, 200 µs probe pulse. Microwaves are for atomic state rotations. A CCD imaging system measures the population difference between the hyperfine states after releasing the atoms from the lattice and spatially separating the states; see text. (**B**) Due to the commensurate frequency relationship between the trapping and the interaction/probe lasers, all atoms are uniformly coupled to the 780 nm mode. (**C**) The 780 nm mode couples the two hyperfine clock states separated by $\omega_{HF}$ to the excited manifold with opposite detunings. Thus, the two states pull the index of refraction seen by the light in opposite directions, leading to a cavity frequency shift proportional to $J_z$. (**D**) The mechanism leading to the collective atomic interactions ($J_z^2$ Hamiltonian) that enables state magnification: linking of the intra-cavity power to $J_z$, producing a $J_z$ dependent ac-Stark shift. The frequencies of the interaction, $\nu_{int}$, and probe $\nu_{probe}$ beams are indicated.

The details of the experimental apparatus are described in refs. *(25, 12)*. We load up to $5\times10^5$ atoms at 25 µK into a 520 µK-deep optical lattice inside the high finesse ($1.75\times10^5$) cavity, and bring the state of the atoms to the equator of the Bloch sphere with a microwave $\pi/2$- pulse. A 780 nm standing-wave cavity mode is used for generating the collective interactions and probing the atoms. The lattice holds the atoms at the intensity maxima of this 780 nm mode, ensuring uniform atom-cavity coupling (Fig. 2A-B). The 1560 nm lattice light whose frequency is stabilized to the cavity generates the 780 nm light to be used through frequency doubling, guaranteeing its frequency stability with respect to the cavity. By measuring the phase of a reflected probe pulse with homodyne detection we can determine the empty cavity frequency down to a $J_z$ equivalent of 3 spin-flips. We use this probe only as a diagnostic tool.

We demonstrate the basic principles of state magnification in Figure 3. To stay within the dynamic range of the cavity measurements ($J_z \sim \pm 800$, equivalent to ~1 cavity linewidth), we limit the atom numbers to $2 \times 10^5$. By applying small microwave rotations (±2 mrad), we alternate between two CSSs with different mean $J_z$ values ($\langle J_z \rangle = \pm 200$). As determined by cavity measurements, the widths of the $J_z$ distributions read within 0.5 dB of the calibrated CSS noise level *(12)*. We then implement the following procedure: excite the cavity with a 200 μs light pulse, detuned by $\delta_0 = 36\,\mathrm{kHz}$, to shear the state; apply a microwave π/2-rotation about axis of the state on the Bloch sphere; and finally release the atoms from the lattice to observe the cloud through fluorescence imaging (2 ms). To spatially separate the two hyperfine states we push the $F = 2$ atoms via resonant absorption. The measured technical noise of 1200 atoms rms of our fluorescence detection setup is ~{15, 10} dB above the SQL for {$2 \times 10^5$, $5 \times 10^5$} atoms.

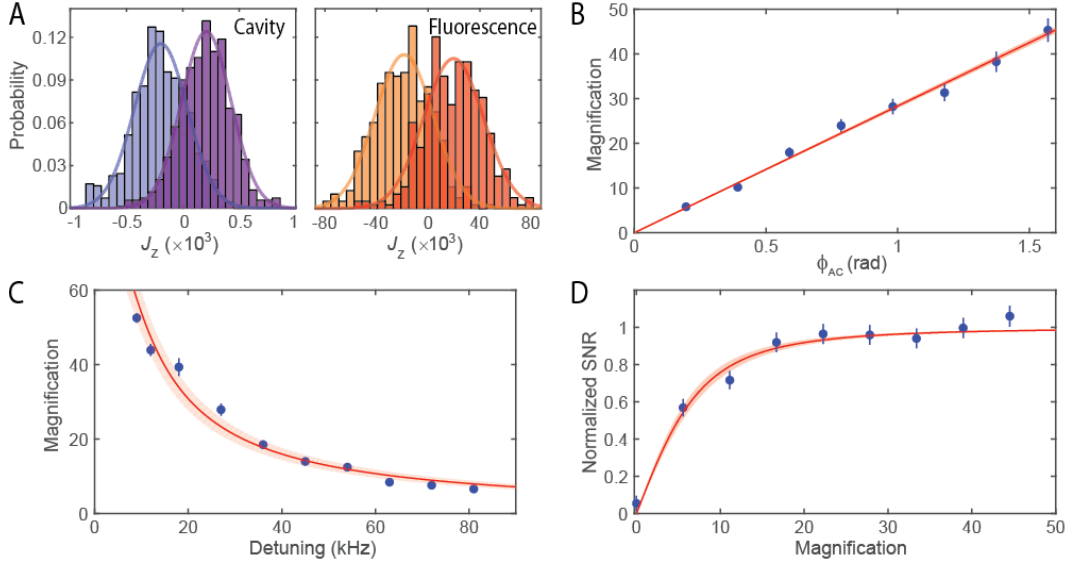

**Fig. 3. Characterization of basic state magnification with CSSs.** (**A**) Sample distributions comparing the cavity based measurements of $J_z$ with fluorescence imaging based measurements after a state magnification of $M = 45$. The two distributions in each plot correspond to different initial states with $\langle J_z \rangle = \pm 200$ prepared using $2 \times 10^5$ atoms. (**B**-**C**) Characterization of the expected dependencies of the magnification factor *M*: Magnification of the separation between the two distributions as a function of accumulated ac-Stark phase shift $\phi_{AC}$ (B) imparted on the atoms at fixed cavity-light detuning of 36 kHz, and as a function of cavity-light detuning (C) at fixed $\phi_{AC} = 0.6\,\text{rad}$. Solid lines: fits to the data as a function of $\phi_{AC}$ and $\delta_0$ respectively in the magnification formula given in text. Fitted curves agree to within 10% with theoretical curves (not shown). (**D**) SNR associated with the two distributions as a function of the magnification parameter, normalized to that obtained by the cavity measurements (Normalized SNR). Magnification is varied by changing $\phi_{AC}$. Solid line: fit of the form $M / (\alpha^2 + M^2)^{1/2}$; fit parameter $\alpha$ contains information primarily about fluorescence detection noise. Error bars and shaded regions in all panels: 68% statistical confidence interval for data and fits respectively.

A comparison of the $J_z$ distributions obtained by cavity and fluorescence measurements indicates a faithful magnification procedure (Fig. 3A). The magnification increases linearly with incident shearing light power (Fig. 3B), quantified by $\phi_{AC}$, and has the expected dependence on cavity-light detuning (Fig. 3C). In the large *M* limit, the signal-to-noise ratio (SNR) associated with the two states after magnification approaches the value measured by the cavity (Fig. 3D), set by the intrinsic sensitivity of the quantum state. Here, SNR is defined as the mean separation of the two states divided by the width of their distributions.

For the mapping to be accurate in this protocol, the magnified $J_z$ noise $M\Delta J_z(0)$ should exceed the initial $J_y$ noise $\Delta J_y(0)$. If we magnify a $J_z$-squeezed state which has $\Delta J_z(0) = \Delta_{CSS}\xi$ and $\Delta J_y(0) = \Delta_{CSS}\xi'$ ($\xi' \cdot \xi \geq 1$, $\xi < 1$), the final noise becomes $\Delta J_y = \left((\Delta_{CSS}\xi')^2 + (M\Delta_{CSS}\xi)^2\right)^{1/2}$. To set the fractional noise contribution of the second term to $1 - \varepsilon$, a magnification of $M_\varepsilon = (2\varepsilon)^{-1/2}\xi' / \xi$ is required. This quantity grows unfavorably (at least quadratically) with the squeezing factor $\xi$.

The unfavorable scaling can be eliminated using a noise re-focusing version of the protocol (illustrated in Figure 1D) which enables, in principle, perfect mapping at a chosen specific magnification. By adding a small rotation $\theta$ before magnification, the $J_y$ noise can be made to focus down through the course of magnification. The action of the small rotation is formally analogous to that of a lens on a beam of light. The final $J_y$ noise in this version of the protocol reads *(23)* $\Delta J_y \approx \left((1 - M\theta)^2(\Delta_{CSS}\xi')^2 + (M\Delta_{CSS}\xi)^2\right)^{1/2}$. For a chosen $\theta = \theta_0$, the initial $J_z$ noise becomes the sole noise contribution at $M = M_0 \equiv 1/\theta_0$.

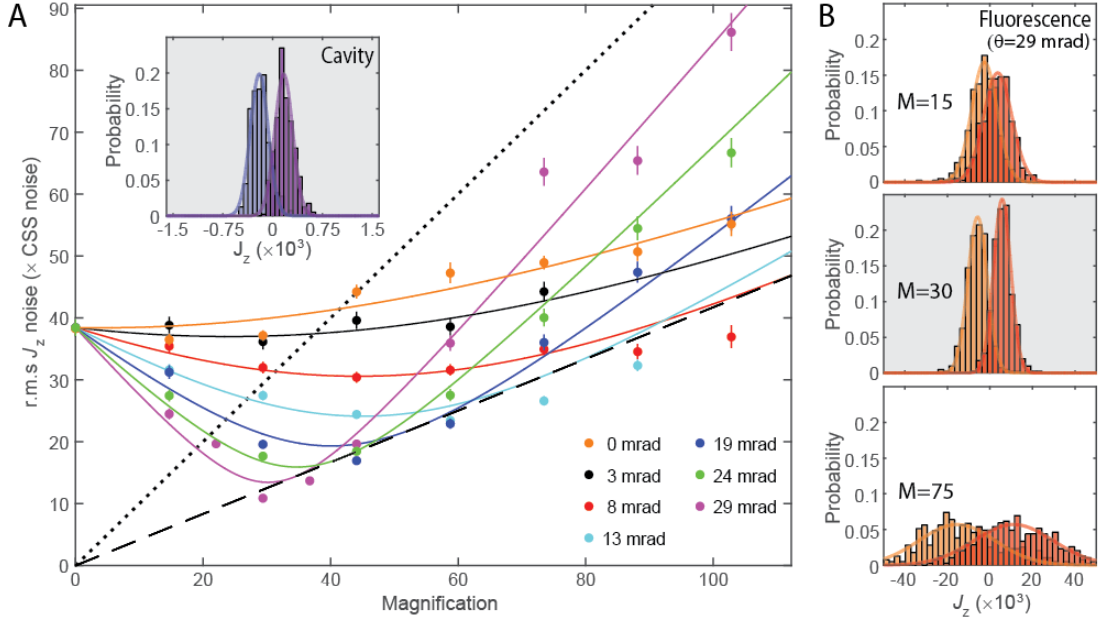

**Fig. 4. State magnification with noise re-focusing using 8 dB squeezed spin-states.** (**A**) Post-magnification $J_z$ noise in units of CSS noise for different amounts of prior on-axis rotation $\theta$ (see Fig. 1D). Solid lines are a global fit to the entire data set with two free parameters: $d\theta/dt$, the rate of change in $\theta$ with microwave pulse time; and the $J_z$ noise of the initial squeezed states. Obtained values are within 15% of the calculated values. The inset shows the distribution of two separated 8 dB squeezed initial states as identified by cavity measurements [to be compared with the $M = 30$ distribution in (B)]. Dashed line: $M$ times the $J_z$ noise contribution from the initial squeezed states. Dotted line: $M$ times the CSS noise – the SQL. Error bars: 68% statistical confidence interval. (**B**) The distributions after the magnification protocol at the indicated $M$ values for $\theta = 29$ mrad. The normalized SNR becomes 0.96±0.06 at $M \approx 1/|\theta|$ (middle histogram), where the dashed line is tangent to the $\theta = 29$ mrad noise curve in (A). For $M$ values to either side of $1/|\theta|$, the two distributions blur into each other (top and bottom histograms).

The noise re-focusing version of the protocol is demonstrated in Figure 4 using squeezed spin states with $5 \times 10^5$ atoms. We generate the states with the same shearing interaction later used for magnification *(1, 23)*. We start with a state 8 dB squeezed in $J_z$ and 32 dB anti-squeezed in $J_y$ ($\xi \approx 0.4$, $\xi' \approx 40$) – the best we can currently achieve without measurement based methods. We apply a set of small microwave rotations $\theta$ about the axis of the state on the Bloch sphere, and investigate the noise measured at the end of the magnification protocol (Fig. 4A). We explicitly demonstrate the existence of an optimal magnification value (here $M \sim 30$) where the magnified state replicates the SNR of the initially prepared states (Fig. 4B). Had we not utilized noise re-focusing, the required magnification would have been $M_{0.05} \sim 320$ for an infidelity $\varepsilon = 0.05$. This level of magnification causes decoherence and starts wrapping the states around the Bloch sphere.

In assessing metrological gain obtained from spin squeezing, the degree of Bloch vector length (coherence) preservation is essential in order not to degrade signal levels. Throughout all state preparation and magnification, the coherence of the states measured by Ramsey fringe contrasts remains above 96%. The small reduction arises from residual atom-cavity coupling inhomogeneities.

We note that this is the first time squeezed states prepared in a cavity are read-out via fluorescence imaging. Here, we focused on quantum metrology without direct measurements below the CSS noise level. However, if used as the read-out stage for the more effective measurement-based squeezing methods, we expect the magnification technique to yield results surpassing the best previous results by improving the read-out *(23)*. Although we only demonstrated state magnification on coherent and squeezed spin states, the method holds generally and can be used to magnify distributions associated with a variety of exotic states. Examples include some states with negative Wigner functions *(16)*, and yet to be demonstrated Schrodinger-cat spin states *(26, 27)*, where cat-fringe spacing can be magnified allowing easier detection. Since the only required key element is a nonlinear phase shift, the method could find broad use in systems utilizing, e.g., collisional interactions in Bose-Einstein condensates *(28, 29,*

*11)*, Rydberg blockade interactions in neutral atoms *(21)*, Ising interactions in ion traps *(30)*, nonlinearities in superconducting Josephson junctions, and nonlinearities in optics. In Supplementary Materials *(23)* we describe a photonic analog of the state magnification concept using self-phase modulation.

**Acknowledgments:** This work was funded by grants from the Defense Threat Reduction Agency and the Office of Naval Research. We thank M. Schleier-Smith for crucial discussions over the course of this work.

**Supplementary Materials:**

<u>Generation</u> of <u>squeezed</u> states

To generate squeezed states, we use the same shearing interaction used for magnification. The manner in which a state becomes squeezed due to the shearing interaction can be seen in Figure 1B. Starting from a CSS, the Wigner distribution becomes narrower in a particular direction as the shearing is increased. To obtain $J_z$-squeezing, a small final microwave rotation about the axis of the state needs to be applied.

The squeezed state generation closely follows the one described in ref. *(1)*. The difference here lies in our use of large cavity-light detunings ( $\delta_0 \sim 4\kappa$ instead of $\delta_0 \sim \kappa/2$ ). We choose a sufficiently large detuning such that the assumed linear intra-cavity power dependence on $J_z$ is valid for the entire CSS noise distribution: At $5 \times 10^5$ atoms, the 4-sigma $J_z$ noise for a CSS is equivalent to $\sim \kappa$ in terms of cavity shift. This is to be compared with our $\sim 4\kappa$ detuning. In this further detuned regime *(31)* the back-action on $J_y$ due to cavity decay (the measurement back-action) is reduced at the expense of a decreased shearing strength for the same amount of input light (Fig. 3C). In principle, this makes the interaction more ideal in terms of noise-area preservation for a given target squeezing level, until the point at which the spin-flips due to spontaneous emission become the limiting factor to the achievable squeezing.

Although the shearing interaction should be near-area preserving for our range of parameters, as is evident from the generated squeezed states with 8 dB squeezing and 32 dB anti-squeezing, there is apparent area non-preservation. This area non-preservation is a result of statistical ensemble averaging in presence of technical noise, effectively giving rise to mixed states. There are two main contributions. The first contribution is from the technical pulse area noise of the light generating the shearing. The $J_y$ noise increases with pulse power due to increasing absolute ac-Stark shift precession noise. The second contribution is from the microwave phase noise. The mean $J_z$ value of the generated squeezed states after the microwave rotation that concludes state preparation is crucially dependent on the location of the rotation axis. For $5 \times 10^5$ atoms, we observe 10-15 dB above CSS noise uncertainty in the rotation axis location due to microwave phase noise. Because of these effects, at low shearing powers we do not observe noise levels below the CSS noise, and at higher levels ( $\phi_{AC} = \pi/8$ ) squeezing saturates around 8 dB.

<u>The</u> <u>shearing</u> <u>interaction</u> in the <u>atom-cavity</u> system

A systematic treatment of the shearing interaction in a driven atom-cavity system is examined in the supplemental materials of ref. *(21)* using the Lindblad master equation formalism. In this section we derive the specific equations used in the main text for the magnified mean and noise expressions. We do this in a targeted fashion using the Heisenberg-Langevin formalism starting from an atom-cavity-reservoir Hamiltonian. We first elucidate the connection between the one-axis twisting Hamiltonian used in the main text and the experimentally realized dispersive atom-cavity system. We then then clarify why the former is sufficient to explain the results of the current work. We use the derived results to analyze the limitations of the magnification scheme in the next section.

We take the interaction picture Hamiltonian for the physical system as

$$H^{(I)} = \hbar \delta_c J_z a^\dagger a - \hbar \delta_c J_z \mid \alpha(0,t) \mid^2 - \hbar \left( \beta^*(t) e^{i\delta_0 t} a + \beta(t) e^{-i\delta_0 t} a^\dagger \right)$$
$$- \hbar \sum_k \sqrt{\tfrac{\kappa c}{2L}} \left( a_k^\dagger a e^{i(\omega_k - \omega_c)t} + a_k a^\dagger e^{-i(\omega_k - \omega_c)t} \right)$$

The first term is the dispersive atom-cavity interaction shifting the cavity resonance frequency by $\delta_c$ per spin-flip, or equivalently generating a differential ac-Stark shift $\delta_c$ on the clock states per intra-cavity photon; $a$ is the cavity mode annihilation operator. The second term is a specific choice of a rotating frame for taking out the mean precession; $\mid \alpha(0,t) \mid^2$ is the time dependent mean intra-cavity photon number for $J_z = 0$. The third term accounts for the excitation of the cavity mode with an incident laser; $\beta(t)$ is the driving amplitude, $\delta_0$ is the cavity-light detuning. The last term phenomenologically induces the decay of the cavity modes at a rate $\kappa$ to a continuum of modes with dispersion relation $\omega_k = ck$ in a 1D-space extending from $-L/2$ to $L/2$ ($L \rightarrow \infty$ limit is taken); $a_k$ are the decay mode annihilation operators, $\omega_c$ is the cavity resonance frequency. Since this decay is phenomenological we also include the loss due to atomic scattering in this term: $\kappa = \kappa_0 + N\delta_c \Gamma / \omega_{HF}$; $\kappa_0$ is the measured empty cavity decay rate, the second term is the total atomic scattering rate assuming $J_z \sim 0$ and atom-cavity detunings of $\pm \omega_{HF} / 2$ as shown in Figure 2C.

The relevant Heisenberg equations of motion that can be derived from the above Hamiltonian using the standard methods of the Heisenberg-Langevin approach *(32)* are

$$J_z(t) = J_z(0)$$
$$\dot{a} = \left( -i\delta_c J_z - \tfrac{\kappa}{2} \right) a + i\beta(t) e^{-i\delta_0 t} + f_a$$
$$\dot{J}_y = -\delta_c \mid \alpha(0,t) \mid^2 J_x + \delta_c a^\dagger J_x a$$

Here $f_a(t) = i \sum_k \sqrt{\tfrac{\kappa c}{2L}} a_k(0) e^{-i(\omega_k - \omega_c)t}$ is a noise operator. This operator and the decay term in the second equation encapsulate the effects of the reservoir.

Now we show that the intra-cavity field amplitude becomes linked to $J_z$ in the limit of slowly varying drive amplitude; $\left| \dot{\beta}(t) / \beta(t) \right| \ll \left| \delta_c J_z - i\kappa / 2 \right|$. We integrate the equation for $\dot{a}$ by parts assuming $\beta(0) = 0$, drop the arising integration involving $\dot{\beta}(t)$, and take the long time limit $t \gg 1/\kappa$ to obtain $a(t) = -\alpha(J_z, t) - f'_a(J_z, t)$. Here $\alpha(J_z, t) = \dfrac{\beta(t) e^{-i\delta_0 t}}{\delta_0 - \delta_c J_z(0) + i\kappa / 2}$, and yields $\alpha(J_z, t) \sim \alpha(0,t)\left(1 + \frac{\delta_c}{\delta_0 + i\kappa/2} J_z\right)$ when Taylor expanded around $J_z = 0$ to obtain the lowest order correction to the field amplitude due to the atoms. $f'_a(J_z, t)$ is a noise operator proportional to $a_k(0)$. Since the decay modes are assumed to be initially in vacuum state, this noise operator will vanish when acting on the initial reservoir state. To facilitate such simplifications we adhere to normal operator ordering.

At this stage we invoke the simplifying assumption that restricts us to a planar patch on the surface of the Bloch sphere, linearizing the equations, with the substitution $J_x \to J \equiv N/2$. This approximation works well if the $J_y$ values remain within $\pm 0.2J$, covering the experimentally relevant range. Substituting the obtained lowest order $a(t)$ into the equation for $\dot{J}_y$ and $dJ_y^2/dt$, and taking the expectation values both over the decay modes and the spins, we get

$$\langle J_y \rangle = \langle J_y(0) \rangle + M \langle J_z(0) \rangle$$

$$\langle J_y^2 \rangle = \langle J_y^2(0) \rangle + M \langle J_z J_y(0) + J_y J_z(0) \rangle + M^2 \langle J_z^2(0) \rangle + \tfrac{N}{4} M \tfrac{\kappa}{\delta_0}$$

Here $M = N \frac{\delta_c \delta_0}{|\delta_0 + i\kappa/2|^2} \phi_{AC}(t)$ with $\phi_{AC}(t) = \int_0^t dt' \delta_c \,|\alpha(0,t)|^2$. The second equation utilizes the intermediate step of obtaining $\langle J_z J_y + J_y J_z \rangle = \langle J_z J_y(0) + J_y J_z(0) \rangle + 2M \langle J_z^2(0) \rangle$ using the first one. These equations allow one to calculate the mean and variance of $J_y$ after magnification in terms of expectation values of initial means and symmetrized (co)variances. We obtain

$$\mathrm{var}(J_y) = \mathrm{var}(J_y(0)) + 2M \,\mathrm{cov}(J_y(0), J_z(0))_{sym} + M^2 \,\mathrm{var}(J_z(0)) + \tfrac{N}{4} M \tfrac{\kappa}{\delta_0}$$

The last term is the only addition to the result that would have been obtained from the one-axis twisting Hamiltonian. It represents the effect of cavity decay (measurement back-action). It is negligible in comparison to the previous terms for our range of parameters, and hence can be dropped. The formula for $\Delta J_y$ in the main text for the noise re-focusing version can be obtained by using the (co)variances appropriate for initially $J_z$-squeezed states rotated by a small angle $\theta$ about the center of the state, and making the small angle and large magnification approximations. The (co)variances are

$$\mathrm{var}(J_y(0)) = \sin^2\theta\,\mathrm{var}(J'_z) + \cos^2\theta\,\mathrm{var}(J'_y)$$

$$\mathrm{var}(J_z(0)) = \cos^2\theta\,\mathrm{var}(J'_z) + \sin^2\theta\,\mathrm{var}(J'_y)$$

$$\mathrm{cov}(J_y(0), J_z(0))_{sym} = -\tfrac{1}{2}\sin 2\theta\left(\mathrm{var}(J'_y) - \mathrm{var}(J'_z)\right)$$

Here the primed variances belong to the $J_z$-squeezed state before rotation. The $\theta = 0$ limit recovers the noise for the basic version of the protocol.

<u>Limits</u>

In ref. *(21)* it has been shown that using the one-axis twisting Hamiltonian $H = \hbar \chi J_z^2$ in the shear-unshear mode, it is in principle possible to reach the Heisenberg limit to measurement precision with a required detection resolution only at the CSS noise level. It was also shown that for an implementation of the sort described in our work, the Heisenberg limit is not achieved due to the atomic and cavity decays, but instead a metrological gain that is limited by $\sim \sqrt{NC}$ is obtained. Here $C$ is the single atom cooperativity. In this section, we analyze the minimum squeezed-state noise that can be resolved using the noise re-focusing magnification protocol that we presented. We build upon the formulas derived in the previous section. We find a similar $\sqrt{NC}$ scaling in the metrological gain with a universal saturation limit of ~30dB for $^{87}$Rb atoms. The important difference is that with the current method, in principle, the detection sensitivity requirement can be relaxed to arbitrarily chosen levels.

We will use the following relations in the derivation: The cooperativity $C$ is defined as $C = 4g^2/\kappa_0 \Gamma \approx 0.78$ with $g$, the atom-cavity coupling; $\Gamma$, the atomic excited state decay rate; $\kappa_0$, the empty cavity decay rate. The atomic-absorption-broadened cavity linewidth is denoted by $\kappa$ with $\kappa/\kappa_0 = 1 + NC(\Gamma/\omega_{HF})^2$. The per spin-flip cavity frequency shift is $\delta_c = 4g^2/\omega_{HF}$.

We assume a given initial $J_z$-squeezed state with $\Delta J_z(0) = \xi \Delta_{CSS} = \xi \sqrt{N}/2$. Working at the optimal magnification point $M = 1/\theta$ ($\theta \ll 1$) of the noise re-focusing scheme, the $J_y$ variance after magnification is given by $\text{var}(J_y) = \frac{N}{4}M^2\xi^2 + \frac{N}{4}M\frac{\kappa}{\delta_0}$. The first term comes purely from the initial $J_z$ noise. The second term is the spurious contribution from the cavity decay, and can be suppressed by increasing the cavity-light detuning $\delta_0$ at the expense of an increased amount of light to achieve the same magnification $M$. Thus, at large detunings one needs to carefully consider the effects of spontaneous emission, which will effectively limit $\delta_0$. As described in ref. *(12)*, spin-flips caused by spontaneous emission events gives rise to a diffusion of $J_z$ with a variance given by $\sigma_{\text{flip}}^2 = \frac{N}{6}\frac{\Gamma}{\omega_{HF}}\phi_{AC} = \frac{1}{6C}M\frac{\delta_0}{\kappa_0}$. The second equality assumes $\delta_0 \gg \kappa/2$.

For the magnification to work properly, by the time one reaches the optimal magnification point, the total spin-flip noise due to spontaneous emission should not exceed the initial $J_z$ noise ($\sigma_{\text{flip}}^2 \leq \xi^2 \frac{N}{4}$). This sets the maximum usable detuning to $\delta_0 = \frac{3}{2}NC\kappa_0\xi^2/M$. Inserting this into the post-magnification $J_y$ variance we obtain: $\text{var}(J_y) = \frac{N}{4}M^2\xi^2 + M^2\frac{1}{\xi^2}\frac{1+NC(\Gamma/\omega_{HF})^2}{6C}$. We will assume that the initial state is no longer accurately resolvable when the second term becomes equal to the first one. Equating the two terms gives us an estimate of the best resolving power for the magnification scheme that our atom-cavity implementation with $^{87}$Rb atoms can achieve in principle. The smallest squeezing parameter that can be resolved is given by

$$\xi_{\min}^2 = \sqrt{\frac{1 + NC(\Gamma/\omega_{HF})^2}{\frac{3}{2}NC}}.$$

It appears that the achievable limits do not depend on the choice of magnification as long as one operates at the optimal point of the noise re-focusing scheme. $\xi_{\min}^2$ first decreases with atom number as $1/\sqrt{\frac{3}{2}NC}$ and saturates at $\xi_{sat}^2 = \sqrt{\frac{2}{3}}(\Gamma/\omega_{HF}) \equiv -31\,\mathrm{dB}$ due to the effects of atomic absorption. For our experimental parameters with $N = 5 \times 10^5$, we obtain $\xi_{\min}^2 \sim -28\,\mathrm{dB}$.

### Other Physical systems

*Optical self-phase modulation* – In this section, we show that the state magnification concept investigated in the context of atomic systems has an optical analogue. We show how the method carries over to optical self-phase modulation in a medium with Kerr nonlinearity. This system magnifies the amplitude of an optical excitation by mapping it onto its phase, which can then be read out with homodyne detection.

The interaction picture Kerr Hamiltonian is given by $H^{(I)} = -\hbar \chi a^{\dagger 2} a^2 + 2\hbar \chi |\alpha|^2 a^{\dagger} a$ *(33)*. Here the first term is the actual Kerr Hamiltonian, and the second term is a choice of a rotating frame (in phase space) that takes out the residual linear phase evolution induced by the first term.

We are interested in initial states with slightly differing mean amplitudes centered about a reference value $|\alpha| \gg 1$, i.e., $(\langle X \rangle^2 + \langle Y \rangle^2)^{1/2} \approx |\alpha|$. Here $X = (a + a^{\dagger})/2$ and $Y = (a - a^{\dagger})/2i$ are the quadrature operators with the commutation relation $[X, Y] = i/2$. To simplify the problem, we go into a displaced picture using the displacement operator $D \equiv D(\alpha)$ (with $\alpha = |\alpha| e^{i\phi}$), whose action can be summarized by the relation $D^{\dagger}(\alpha) a D(\alpha) = a + \alpha$. This takes out the large quantity $|\alpha|$ from the states, and makes it explicitly a part of the new Hamiltonian $H^{(DI)} = D^{\dagger} H^{(I)} D$. The displaced-picture quantum states are related to the original picture by $|\psi^{(D)}\rangle = D^{\dagger} |\psi\rangle$. In terms of the displaced-picture operators $\tilde{X}$, $\tilde{Y}$ and $\tilde{a}$, the original operators are $X \equiv \tilde{X} + \alpha_{Re}$, $Y \equiv \tilde{Y} + \alpha_{Im}$ and $a \equiv \tilde{a} + \alpha$.

Ignoring the small terms proportional to $|\alpha|^1$ and $|\alpha|^0$, and dropping the constant terms as they arise, the displaced-picture Hamiltonian is given by

$$H^{(DI)} \approx -\hbar \chi \left( \alpha^2 \tilde{a}^{\dagger 2} + \alpha^{*2} \tilde{a}^2 + 2|\alpha|^2 \tilde{a}^{\dagger} \tilde{a} \right) = -\hbar 4\chi |\alpha|^2 (\cos\phi \tilde{X} + \sin\phi \tilde{Y})^2$$

For simplicity, we assume that the states of interest are near the *X*-axis in phase space, and thus we set $\phi = 0$, $\alpha_{Im} = 0$, and obtain $H^{(DI)} = -4\hbar \chi |\alpha|^2 \tilde{X}^2$. The Heisenberg equations of motions are given by

$$\dot{\tilde{X}} = 0$$

$$\dot{\tilde{Y}} = 4\chi \, |\, \alpha\, |^2 \, \tilde{X}$$

Solving these equations, for the original observables we obtain

$$\langle X(t) \rangle = \langle \tilde{X}(0) \rangle + \alpha_{\text{Re}}$$

$$\langle Y(t) \rangle = \langle \tilde{Y}(0) \rangle + M \langle \tilde{X}(0) \rangle$$

$$\text{var}(X(t)) = \text{var}(\tilde{X}(0))$$

$$\text{var}(Y(t)) = \text{var}(\tilde{Y}(0)) + 2M \, \text{cov}(\tilde{X}(0), \tilde{Y}(0))_{\text{sym}} + M^2 \, \text{var}(\tilde{X}(0))$$

Here $M = 4\chi \, |\, \alpha\, |^2 \, t$, and expectation values are taken using the displaced-picture states. These equations are directly analogous to the shearing interaction we implement for our atomic system: The initial amplitude quadrature signal is mapped onto the phase quadrature with magnification $M$, which can be detected with homodyne detection. With appropriate initial states, containing correlations between the two quadratures, i.e., $\text{cov}(\tilde{X}(0), \tilde{Y}(0))_{\text{sym}} \neq 0$, noise re-focusing schemes are possible here as well.

The analysis provided here is valid as long as the mean $X$ values are much larger than the values associated with the post-magnification $Y$ distributions. Failure to satisfy this condition makes the approximation for the simplified displaced-picture Hamiltonian invalid. Outside of the regime considered here, the states are driven into crescent shapes in phase space *(33)*.